%% file: main.tex
\documentclass[letterpaper, 10 pt, conference]{ieeeconf}  

\IEEEoverridecommandlockouts                              

\usepackage{graphics} 
\usepackage{epsfig} 
\usepackage{mathptmx} 
\usepackage{times} 
\usepackage{amsmath} 
\usepackage{amssymb}  
\usepackage{booktabs}       
\usepackage{xcolor}         
\usepackage{soul}           
\usepackage{multirow}
\usepackage{hyperref}
\usepackage{xurl}
\usepackage{mathrsfs}
\usepackage{svg}
\usepackage[colorinlistoftodos]{todonotes}

\title{\LARGE \bf
Analyzing Emissions and Energy Efficiency at Unsignalized Real-world Intersections Under Mixed Traffic Control 
}

\author{Michael Villarreal, Dawei Wang, Jia Pan, Weizi Li
\thanks{Michael Villarreal and Weizi Li are with Min H. Kao Department of Electrical Engineering and Computer Science, University of Tennessee, Knoxville, Knoxville, TN 37996, USA. 
{\tt\small tvillarr@vols.utk.edu, weizili@utk.edu}}%
\thanks{Dawei Wang and Jia Pan are with Department of Computer Science at the University of Hong Kong.
{\tt\small dawei@connect.hku.hk, jpan@cs.hku.hk}}%
}

\begin{document}

\maketitle
\thispagestyle{empty}
\pagestyle{empty}


\begin{abstract}
Greenhouse gas emissions have dramatically risen since the early 1900s with U.S. transportation generating 28\% of U.S. emissions. As such, there is interest in reducing transportation-related emissions. Specifically, sustainability research has sprouted around signalized intersections as intersections allow different streams of traffic to cross and change directions. Recent research has developed mixed traffic control eco-driving strategies at signalized intersections to decrease emissions. However, the inherent structure of a signalized intersection generates increased emissions by creating frequent acceleration/deceleration events, excessive idling from traffic congestion, and stop-and-go waves. Thus, we believe unsignalized intersections hold potential for further sustainability improvements. In this work, we provide an emissions analysis on unsignalized intersections with complex, real-world topologies and traffic demands where mixed traffic control strategies are employed by robot vehicles (RVs) to reduce wait times and congestion. We find with at least 10\% RV penetration rate, RVs generate less fuel consumption, CO$_2$ emissions, and NOx emissions than signalized intersections by up to 27\%, 27\% and 28\%, respectively. With at least 30\% RVs, CO and HC emissions are reduced by up to 42\% and 43\%, respectively. Additionally, RVs can reduce network-wide emissions despite only employing their strategies at intersections. 

\end{abstract}

\input{sections/intro}
\input{sections/related}

\input{sections/methodology} 
\input{sections/results}
\input{sections/conclusion}
\input{sections/acknowledge}







\bibliographystyle{unsrt}
\bibliography{ref}

\end{document}

%% file: sections/intro.tex
\section{INTRODUCTION}

Greenhouse gas emissions have steadily risen since the early 1900s with a 90\% increase 
from the 1970s to current day~\cite{epa2023emissions}. Increased emissions is monumentally detrimental to our planet's health causing increased surface temperatures, irregular precipitation patterns and storm severities, and rising sea levels~\cite{eia2023effects}. A major contributing factor to increased emissions is the transportation sector. For example, transportation is responsible for 28\% of all United States greenhouse gas emissions~\cite{ghg2023energy}. Thus, there is a particular interest in studying effective ways for reducing transportation-related emissions. 
In particular, a widely-prevalent transportation infrastructure that has seen sustainability research given its crucial role is the intersection~\cite{brian2009stochastic, aziz2018learning, xie2014real, ge2013energy, jayawardana2022learning, bai2022hybrid}.

Intersections are vital components to our transportation infrastructure as intersections allow traffic flowing in different directions to interchange and disperse. A typical intersection is signalized where traffic signal phase patterns control traffic flows. However, from a sustainability standpoint, signalized intersections are often insufficient as they inherently increase generated emissions~\cite{barth2008real}.
Signalized intersections generate increased emissions from stop-and-go waves forming due to traffic signal patterns as vehicles experience increased frequencies of severe acceleration and deceleration events. Additionally, traffic congestion, which causes excessive idling and prevents vehicles traveling at optimal speeds for reduced emissions, can result from traffic signals~\cite{barth2008real}.
One avenue for improved sustainability at intersections is sustainable traffic signal control strategies~\cite{brian2009stochastic, aziz2018learning, xie2014real, ge2013energy}; however, this restricts the domain to signalized intersections and its associated pitfalls. 

Another avenue is eco-driving strategies at intersections where robot vehicles are used to control mixed traffic to reduce emissions production. 
Recent works, such as by Jayawardana et al.~\cite{jayawardana2022learning} or Bai et al.~\cite{bai2022hybrid}, study eco-driving strategies at signalized intersections through mixed traffic control. Mixed traffic control involves using robot vehicles (RVs) to regulate human-driven vehicles (HVs) to improve traffic conditions. The RVs are trained using reinforcement learning (RL) as RL can handle complex mixed traffic scenarios without using predefined models or heuristics. These works leverage RL's ability to optimize multiple objectives to achieve reduced emissions while minimally impacting vehicle travel times. However, signalized intersections are still used meaning the aforementioned pitfalls still exist, even if in a reduced form. Thus, a recent approach to intersection traffic control that aims to avoid signalized intersection pitfalls is mixed traffic control via RL at unsignalized intersections.

\begin{figure}
    \centering
    \includegraphics[width=1.0\linewidth]{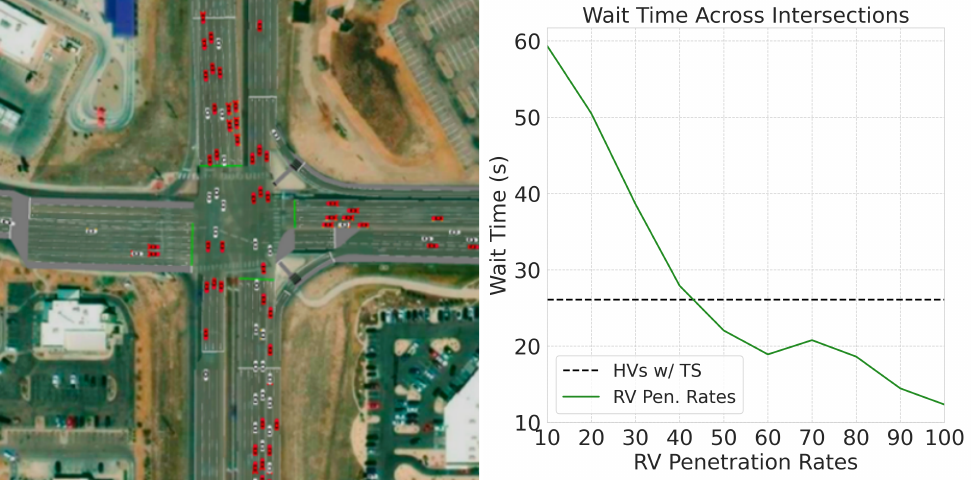}
    \caption{LEFT: One of the complex real-world unsignalized intersections Wang et al.~\cite{wang2023learning} train on with mixed traffic (RV in red and HV in white). 
    RIGHT: Wang et al. show RVs significantly reduce wait times compared to HVs with signalized intersections past 60\% RV penetration rate. Wait time results are averaged across four unsignalized intersections.}
    \label{fig:efficiency}
    \vspace{-1.5em}
\end{figure}

Wang et al.~\cite{wang2023learning} explore using mixed traffic control via RL on large, complex, real-world unsignalized intersections. The authors train on four intersections with real-world traffic demand from Colorado, CO, USA (Fig.~\ref{fig:efficiency} LEFT shows one of the four intersections). Wang et al.~\cite{wang2023learning} demonstrate how RVs at unsignalized intersections can provide safety and dramatically reduce average vehicle wait times at RV penetration rates higher than 60\% (shown in Fig.~\ref{fig:efficiency} RIGHT). Their work first showcases the feasibility of controlling mixed traffic at unsignalized intersections with complex topologies and real-world traffic demands. As such, this presents an opportunity to analyze the current state of mixed traffic control strategies at real-world unsignalized intersections to investigate their potential as eco-driving strategies. Given the RVs reduce wait times for vehicles while effectively reducing stop-and-go wave formation, we theorize reduced emissions and improved sustainability will follow. 

In this work, we perform an emissions analysis on fuel, CO$_2$, CO, HC, and NOx emissions at four real-world unsignalized intersections. We compare performance of RVs (at 10\% RV penetration rate increments from the range [10\%, 100\%]) to only HVs at the same intersections, but with traffic signals. We also examine acceleration profiles of different RV penetration rates to gain insights, and we examine the potential effects mixed traffic control strategies have at unsignalized intersections on the emissions generated for the entire road network. Our observations are the following. 
\begin{itemize}
    \item For fuel consumption, CO$_2$ emissions, and NOx emissions, every RV penetration rate outperforms generated emissions compared to HVs with signalized intersections. We see performance improvements by up 27\% for fuel consumption and CO$_2$ emissions and 28\% for NOx emissions. For CO and HC emissions, at least 30\% RV penetration rate is required to outperform HVs with signalized intersections with improvements up to 42\% and 43\%, respectively.
    \item When examining fuel consumption at the individual intersections, we observe RVs with at least 30\% RV penetration rate outperform HVs with signalized intersections. However, we find for 10\% and 20\%, average fuel consumption is off only by 2\% from the baseline.
    \item RVs at lower RV penetration rates implement control strategies resulting in average vehicle accelerations stabilizing near 0, meaning less emissions are generated compared to more severe acceleration values.
    \item For emissions across the whole road network, RVs outperform HVs with signalized intersections at every RV penetration rate for fuel consumption and CO$_2$, at least 60\% RV penetration rate for CO and HC emissions, and at least 20\% RV penetration rate for NOx emissions.
\end{itemize}
To the best of our knowledge, we are the first to conduct an emissions analysis on unsignalized intersections where RVs employ mixed traffic control strategies on mixed traffic.

%% file: sections/related.tex
\section{RELATED WORK}

Many studies research emissions generated at intersections. For example, Pandian et al.~\cite{pandian2009evaluating} give an in-depth look at the traffic, vehicle, and road factors affecting emissions at intersections. Other research studies the emission and dispersion factors behind vehicle pollutants at intersections~\cite{goel2014review}, how reducing traffic demand or lowering a road's speed limit can reduce emissions~\cite{mahmod2013reducing}, or how cruising and acceleration can generate 80\% of emissions produced~\cite{papson2012analysis}. Other research investigate how roundabouts (an unsignalized intersection control infrastructure) can reduce certain emissions compared to signalized intersections~\cite{fernandes2017assessing, meneguzzer2017comparison}. 

Other studies research the effect of mixed traffic on generated emissions. Yao et al.~\cite{yao2023analysis} examine the effect of platoon size of connected autonomous vehicles (CAVs; or RVs in this work) on emissions and finds a greater platoon size can lead to greater generated emissions. Other studies have found the inclusion of CAVs can lead to a significant reduction in generated emissions with 100\% RV penetration rate bringing the greatest reduction~\cite{zhao2022fuel, yao2021fuel}.

Numerous studies research different eco-driving strategies for vehicles at signalized intersections, whether by traditional optimization techniques or reinforcement learning (RL) techniques~\cite{wang2019cooperative, lin2020eco, rakha2011eco, chen2019gap}. However, these studies only consider the positive impact of these eco-driving strategies on emissions and do not consider the potential negative impact on vehicle travel time. Works by Jayawardana et al.~\cite{jayawardana2022learning} or Bai et al.~\cite{bai2022hybrid} consider both when developing their eco-driving strategies. Our emissions analysis investigates the impact of emissions, but we consider vehicle wait time instead of vehicle travel time.
However, all mentioned works use signalized intersections, which come with potential pitfalls such as increased wait times, higher congestion levels, and more frequent stop-and-go waves. Our analysis focuses solely on unsignalized intersections to avoid these inherent pitfalls of signalized intersections.

The works by Jayawardana et al.~\cite{jayawardana2022learning} and Bai et al.~\cite{bai2022hybrid} both use RL to develop their eco-driving mixed traffic control strategies. These works demonstrate RL's effectiveness and suitably for mixed traffic control tasks with multiple objectives. RL has previously been shown to be well-suited for mixed traffic control with other tasks, such as stabilizing a ring road~\cite{wu2021flow} or improving traffic flow in a bottleneck or merge road network~\cite{vinitsky2018benchmarks}.

%% file: sections/methodology.tex
\section{METHODOLOGY}
In the following, we introduce mixed traffic control, the formulation of mixed traffic control as reinforcement learning (RL) tasks, and discuss the unsignalized intersection details~\cite{wang2023learning}.

\subsection{Mixed Traffic Control as Multi-agent Reinforcement Learning}

Mixed traffic control is using robot vehicles (RVs) to control human-driven vehicles (HVs) to improve traffic conditions through learned strategies. In this work, the specific condition being improved is wait time of vehicles at the unsignalized intersections with the goal being minimization. We explore if the learned strategies for wait time minimization also act as eco-driving strategies reducing generated emissions. The RVs are trained and learn mixed traffic control strategies through multi-agent reinforcement learning.

We model mixed traffic control as a Partially Observable Markov Decision Process (POMDP) represented by a tuple ($S$, $A$, $P$, $R$, $p_0$, $\gamma$, $T$, $\Omega$, $O$) where $S$ is the state space; $A$ is the action space; $P(s'|s,a)$ is the transition probability function; $R$ is the reward function; $p_0$ is the initial state distribution; $\gamma\in(0, 1]$ is the discount factor; $T$ is the episode length (horizon); $\Omega$ is the observation space; and $O$ is the probability distribution of retrieving an observation $\omega \in \Omega$ from a state $s \in S$. 
At each timestep $t \in [1,T]$, a RV uses its policy $\pi_{\theta}(a_t|s_t)$ to take an action $a_t$ $\in$ $A$, given the state $s_t$ $\in$ $S$. The RV's environment provides feedback from taking action $a_t$ by calculating a reward $r_t$ and transitioning the agent into the next state $s_{t+1}$. The RV's goal is to learn a policy $\pi_{\theta}$ that maximizes the discounted sum of rewards, i.e., return, $R_t = \sum^{T}_{i=t}\gamma^{i-t}r_i$. Rainbow Deep Q Learning ~\cite{hessel2018rainbow} is used to learn $\pi_{\theta}$. Our environment is multi-agent with each RV using a shared policy to take individual actions.

\subsection{Mixed Traffic Control At Unsignalized Intersections} 

\begin{figure}
    \centering
    \includegraphics[width=\linewidth]{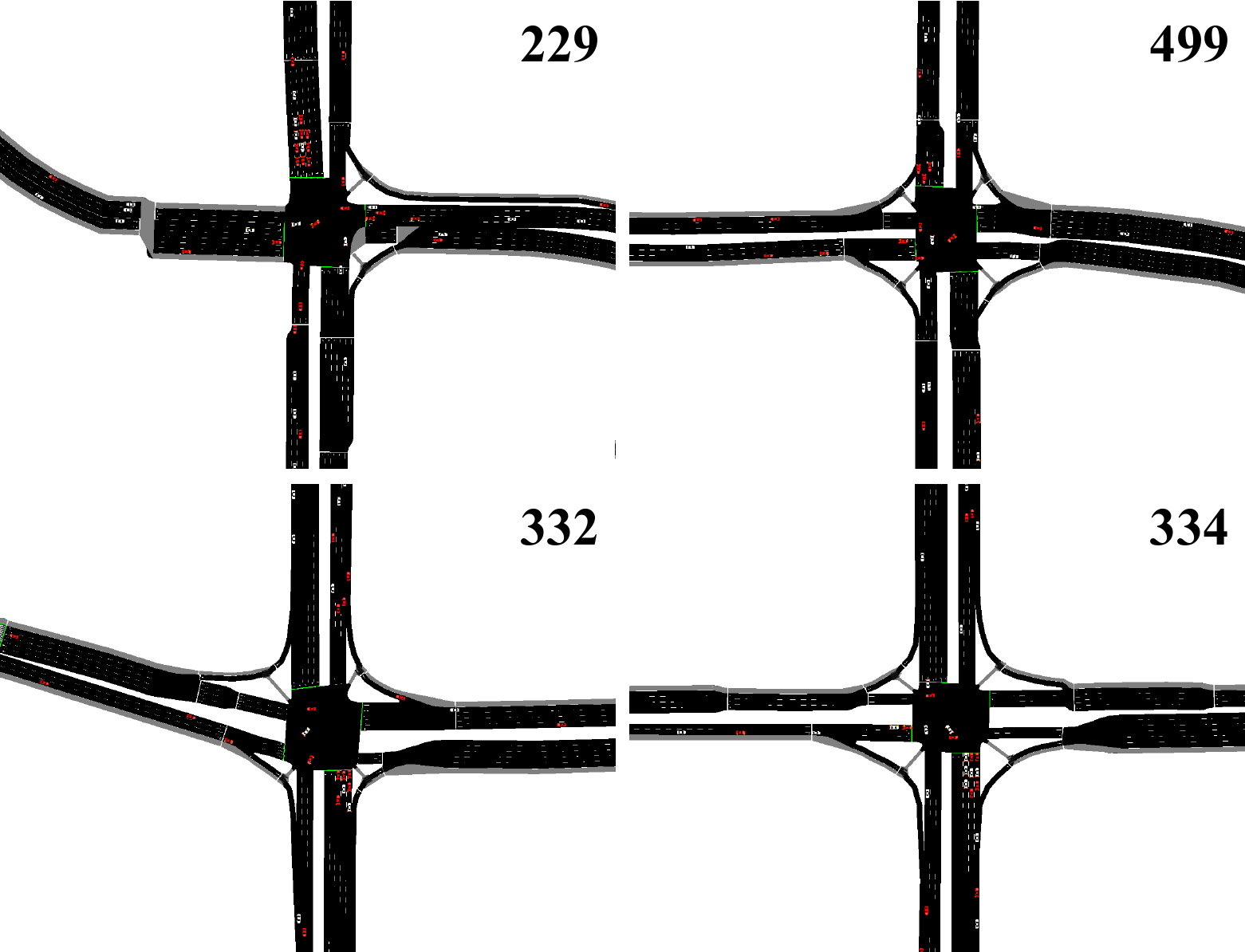}
    \caption{We perform our emissions analysis on four, four-way intersections originating from Colorado Springs, CO, USA. Each of the intersections has been accurately recreated in SUMO~\cite{behrisch2011sumo}. We label the intersections as 229, 499, 332, and 334, from north to south geographically. The traffic flows for each intersection come from real-world turning count data, meaning they represent real-world traffic demand on complex intersections~\cite{wang2023learning}. The RVs (red; human-driven vehicles are white) learn to control and coordinate mixed traffic in the intersection to reduce wait times, while preventing any conflicts. The intersections are learned on individually; however, as the intersections simultaneously exist during learning, what the RVs do at one intersection can affect the rest. We investigate if the learned mixed traffic control strategies also reduce emissions at the intersections.}
    \label{fig:sumo_ints}
    \vspace{-1.5em}
\end{figure}

\subsubsection{Overview} 
We train the RVs at four, four-way unsignalized intersections, shown in Fig.~\ref{fig:sumo_ints}, that originate from Colorado Springs, CO, USA. We label the intersections $229$, $499$, $332$, $334$ from north to south geographically. The real-world intersections are reconstructed in Simulation of Urban Mobility (SUMO)~\cite{behrisch2011sumo} using turning count data to reconstruct the traffic demand~\cite{wang2023learning}. Table~\ref{tab:int_details} illustrates more details. 
Each intersection has at least $16$ lanes, allowing a high amount of through traffic. When recording traffic demand, not every lane experienced traffic so the number of non-empty lanes is, from north to south in Fig.~\ref{fig:sumo_ints}, is 19, 18, 17, and 14, respectively. Total number of lanes for each intersection is given in Table~\ref{tab:int_details}.
The traffic demand for each intersection for each lane, from north to south, is $1157$ vehicles/hour (veh/hr), $1089$ veh/hr, $928$ veh/hr, $789$ veh/hr. The four intersections are reconstructed into the same SUMO network; however, the training occurs separately at each intersection.

\renewcommand{\arraystretch}{1.1}
\begin{table}
    \centering
    \begin{tabular}{ccc}
         \toprule
         Intersection & Num. Incoming Lanes & Traffic Demand \\
         \midrule
         229 & 21 & 1157 \\
         499 & 19 & 1089 \\
         332 & 18 & 928 \\
         334 & 16 & 789 \\
         \bottomrule
    \end{tabular}
    \caption{Details for each of the unsignalized intersections. Traffic demand is measured in vehicles/hour for each lane.}
    \label{tab:int_details}
    \vspace{-1.5em}
\end{table}

The directions the vehicles can travel in are restricted to the Cartesian product of \{North, East, South, West\} x \{Straight, Left Turn\}. Right-turning traffic is not considered for several reasons: 1) most large intersections in the U.S. have dedicated right-turn lanes, 2) traffic laws allow right-turning traffic to turn without a green signal, and 3) Wang et al.~\cite{wang2023learning} demonstrate this decision has minimal impact on controlling and coordinating traffic at intersections. As there are conflicting directions vehicles travel in, the RVs' reward function takes into account conflicts when calculating the reward. Specifically, the direction pairs without conflict are: \{(N-S, N-L), (E-S, E-L), (E-L, W-L), (S-S, N-S), (S-L, N-L), (S-S, S-L), (W-S, E-S), (W-C, W-L)\}.

RVs at each unsignalized intersection only take actions and make observations within a control zone, which begins $30$ meters from the entrance to the intersection. When RVs are outside of the control zone, they act as HVs using the Intelligent Driver Model~\cite{treiber2000congested} for microscopic behavior. The Intelligent Driver Model and the underlying real-world traffic modeling (as no collisions were recorded when collecting traffic data) allow for safety of HVs even when RV penetration rates are low.


\subsubsection{Action Space}
The RVs take a high-level action $a$ from \{Stop, Go\}. Stop indicates the RV will not enter the intersection, stopping any following vehicles. Go indicates the RV will enter the intersection to complete its direction. The RVs are trained not to take the Go action should taking this action cause a conflict within the intersection.

\subsubsection{Observation Space}
The observation space for the RVs begin $30$~meters from each intersection. The observation space at each timestep for a RV is
\begin{equation}
    o = \oplus^{D}_{d} \: \{q_{d}, w_{d}\} \:  \oplus^{D}_{d} \: m_{d}, 
\end{equation}
where $D$ is the set of all directions, $d$ is an individual direction, $q$ is the queue length of RVs, $w$ is the average wait time of the RVs, and $m$ is an occupancy map within the intersection. The occupancy map is a grid, positioned inside the intersection, where grid spaces are marked if a vehicle is in that particular grid space.

\begin{figure*}
    \centering
    \includegraphics[width=1.0\linewidth]{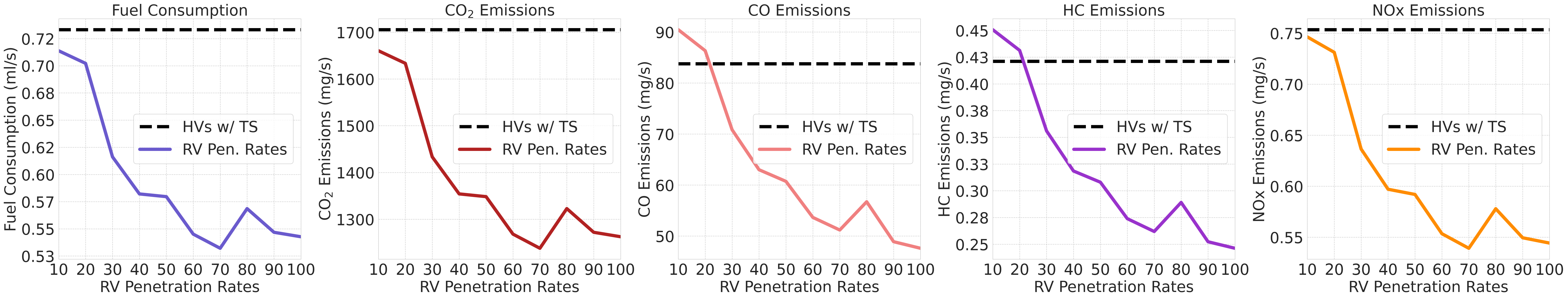}
    \caption{Results for fuel consumption and CO$_2$, CO, HC, and NOx emissions averaged across the four unsignalized intersections. We measure in 10\% increments from 10\% to 100\% RV penetration rate. The dotted, black line represents the baseline HVs with signalized intersections performance. For fuel consumption, CO$_2$ emissions, and NOx emissions, every RV penetration rate outperforms the baseline by up to 27\%, 27\%, and 28\% (at 70\% RV penetration rate), respectively. CO and HC emissions requires at least 30\% RV penetration rate to outperform the baseline. The largest improvement for CO and HC emissions is 43\% and 42\% (both at 100\% RV penetration rate), respectively. We observe that despite not outperforming HVs with signalized intersections on wait time until 50\% (see Fig.~\ref{fig:efficiency}), we are still able to reduce generated emissions at lower RV penetration rates. Overall, the RVs employ mixed traffic control strategies that naturally reduce generated emissions despite not being trained to intentionally reduce emissions.}
    \label{fig:avg_int_results}
    \vspace{-1.5em}
\end{figure*}

\subsubsection{Reward Function}
The reward function encourages individual RVs to lower wait times for other vehicles and consider potential conflicts from entering the intersection. The reward function at each timestep is
\begin{equation}
    r = r_{L} + p_{c}, 
\end{equation}
where $r_{L}$ is:
\begin{equation}
    r_{L} = 
    \begin{cases}
        -w_{d}, & \text{if } a~\text{= Stop;}\\
        w_{d},              & \text{otherwise}
    \end{cases}
\end{equation}
where $w_{d}$ is the average wait time of all vehicles going in direction $d$, normalized in [0,1]. The equation for $p_{c}$ is:
\begin{equation}
    p_{c} = 
    \begin{cases}
        -1, & \text{if conflict;}\\
        0,  & \text{otherwise}
    \end{cases}
    \label{reward_function}
\end{equation}

where $p_{c}$ acts as a penalty term if the RVs movement causes a conflict with other vehicles in the intersection.

\subsection{Analysis of Emissions}

We collect emissions data using SUMO's built-in HBEFA3-based emission model for each vehicle. The emissions are modeled by the equation:
\begin{equation}
    \text{e} = max((\text{f$_1$} + \text{f$_2$}av + \text{f$_3$}a^2v + \text{f$_4$}v + \text{f$_5$}v^2 + \text{f$_6$}v^3) / \text{s} ,0), 
\end{equation}
where each f$_\text{i}$ is an emissions constant associated with the vehicle emission class and emission type (such as fuel or CO$_2$); $a$ is the vehicle's current acceleration; $v$ is the vehicle's current velocity; and $s = 2671.2$, which is a scaling factor. 
The RVs and HVs are passenger vehicles with the ``PC\_G\_EU4" emission class as defined by SUMO, which means the vehicles are gasoline-powered passenger vehicles according to Euro norm four.
The emissions model provides emissions data for several pollutants; as such, we consider fuel consumption, CO$_2$ (carbon dioxide), CO (carbon monoxide), HC (hydrocarbons), and NOx (nitrous oxides) emissions due to their relevancy in sustainability and climate change discussions. 
While not greenhouse gases, we consider fuel consumption given it produces greenhouse gases and CO emissions as CO indirectly aids with climate change~\cite{co2023emissions}.

To conduct the analysis, we train RL policies in 10\% RV penetration rate increments from 10\% to 100\%. For each RV penetration rate, we select the trained policy with the minimum average wait time. We evaluate the selected policies $10$ times and report the averaged emissions results over the last $500$ seconds of a $1000$ second evaluation. 
We report average emissions to make the results independent of the number of cars passing through the intersections given this number can change each timestep/second for any given intersection.
We compare the policies to a baseline of only HVs at signalized intersections/with traffic signals, i.e., 0\% RV penetration rate.

%% file: sections/results.tex
\section{RESULTS}
In this section, we outline the experiment setup and then present the results with discussion of our findings.

\subsection{Experiment Setup}

We train each policy for 1000 iterations using the Rainbow DQN~\cite{hessel2018rainbow} reinforcement learning algorithm. A policy is a neural network of three hidden layers, each with $512$ hidden units. We use a learning rate of $0.0005$ and discount factor of $0.99$~\cite{wang2023learning}. The policies are trained on an Intel i9-13900K CPU with a NVIDIA Geforce RTX 4080 GPU.

\subsection{Emissions Analysis} 

\begin{table*}
    \centering
    \begin{tabular}{lccccccccccc}
        \toprule
        & \multicolumn{11}{c}{Fuel Consumption (ml/s)} \\
         \cmidrule(lr){2-12} 
         & & \multicolumn{10}{c}{RV Penetration Rate} \\
         \cmidrule(lr){3-12}
         & HVs w/ TS & 10\% & 20\% & \textbf{30\%} & \textbf{40\%} & \textbf{50\%} & \textbf{60\%} & \textbf{70\%} & \textbf{80\%} & \textbf{90\%} & \textbf{100\%} \\
         \midrule
         Intersection 229 & 0.7393 & 0.7536 & 0.7541 & 0.6305 & 0.5946 & 0.5773 & 0.5451 & 0.5104 & 0.5513 & 0.5314 & 0.5064 \\
         Intersection 499 & 0.7537 & 0.6809 & 0.6745 & 0.6097 & 0.5822 & 0.5767 & 0.5535 & 0.5364 & 0.5704 & 0.5622 & 0.5483 \\
         Intersection 332 & 0.7298 & 0.7416 & 0.7150 & 0.6341 & 0.5945 & 0.6011 & 0.5581 & 0.5583 & 0.5925 & 0.5615 & 0.5744 \\
         Intersection 334 & 0.7099 & 0.6791 & 0.6654 & 0.5903 & 0.5573 & 0.5634 & 0.5261 & 0.5233 & 0.5602 & 0.5320 & 0.5418 \\
         \bottomrule
    \end{tabular}
    \caption{Fuel consumption results for each of the intersections for the baseline HVs with traffic signals and RV penetration rates. Bold penetration rates mean the RVs generated less emissions than HVs at signalized intersections for every intersection. With at least 30\% RV penetration rate, the RVs employ mixed traffic control strategies leading to reduced emissions at the four unsignalized intersections. At 20\% RV penetration rate, the RVs slightly generate more emissions than the baseline at Intersections 229 by 2\%. HVs with signalized intersections outperform RVs at 10\% penetration rate by 2\% at Intersections 229 and 332. Given these percentages, we believe training the RVs to balance efficiency and emissions could result in the RVs outperform HVs with signalized intersections at every unsignalized intersection.}
    \label{tab:inter_fuel}
    \vspace{-1.5em}
\end{table*}

\subsubsection{Intersection Results}

\setlength{\tabcolsep}{0.5em} 

Fig.~\ref{fig:avg_int_results} presents our results when averaging fuel (ml/s) and CO$_2$ (mg/s), CO (mg/s), HC (mg/s), and NOx (mg/s) emissions across the intersections for each RV penetration rate and HVs with signalized intersections/traffic signals. Each emission type is on the y-axis, while the RV penetration rates are the x-axis.

For fuel consumption and CO$_2$/NOx emissions, the RVs generate less emissions compared to the baseline HVs with signalized intersections at every RV penetration rate. The biggest improvements over the baseline are attained at 70\% RV penetration rate by 27\% for fuel and CO$_2$ and 28\% for NOx. This means that without specifically training to reduce these emissions, only 10\% RV penetration rate is required to reduce emissions and outperform HVs with signalized intersections. For CO and HC emissions, at least 30\% RV penetration rate is required to outperform the baseline by up to 42\% and 43\%, respectively. These significant performance improvements are obtained at 100\% RV penetration rate, but we note significant improvements to generated emissions with at least 30\% RV penetration rate by 15\% for both CO and HC emissions. Overall, the RVs naturally employ mixed traffic control strategies that significantly reduce generated emissions by up to 43\% without explicit training.

For the RV penetration rates that reduce wait time at the intersections, it is understandable for their generated emissions to be less given there is less congestion and excessive idling and potentially less stop-and-go waves. However, we note the magnitude at which emissions are reduced at certain RV penetration rates is not proportional to the reduction in wait time. For example, at 50\% RV penetration rate, wait time is reduced by 15\%, but fuel consumption is reduced by 21\%. Combining this observation with the 10\% to 40\% RV penetration rates reducing emissions without improving wait time reiterates the RVs are employing mixed traffic control strategies that naturally lead to reduced emissions. The worst performing RV penetration rate, 10\%, is off by 7\% from HVs with signalized intersections on CO and HC emissions; this means with improved training, every RV penetration rate outperforming HVs at signalized intersections is feasible.

In Table~\ref{tab:inter_fuel}, we breakdown fuel consumption at the individual unsignalized intersections for the RV penetration rates and HVs with signalized intersections. We consider fuel consumption as an example for the other emissions. Bold RV penetration rates indicate the RVs outperform HVs with signalized intersection at every intersection.

In contrast to Fig.~\ref{fig:avg_int_results}'s results, at least 30\% RV penetration rate is necessary to outperform the baseline at every unsignalized intersection. At 20\% RV penetration rate, the RVs slightly underperform HVs at signalized intersections at Intersection 229 by 2\%. At 10\% RV penetration rate, HVs with signalized intersections outperform RVs at Intersections 229 and 332 by 2\%. Given these close percentages, we believe with improved training, designed specifically to leverage generated emissions in the reward function, every RV penetration rate will outperform the baseline on all four intersections. Overall, RVs with at least 30\% RV penetration rate outperform HVs at signalized intersections at each unsignalized intersection.

\subsubsection{Acceleration Profiles}


\begin{figure*}
    \centering
    \includegraphics[width=\linewidth]{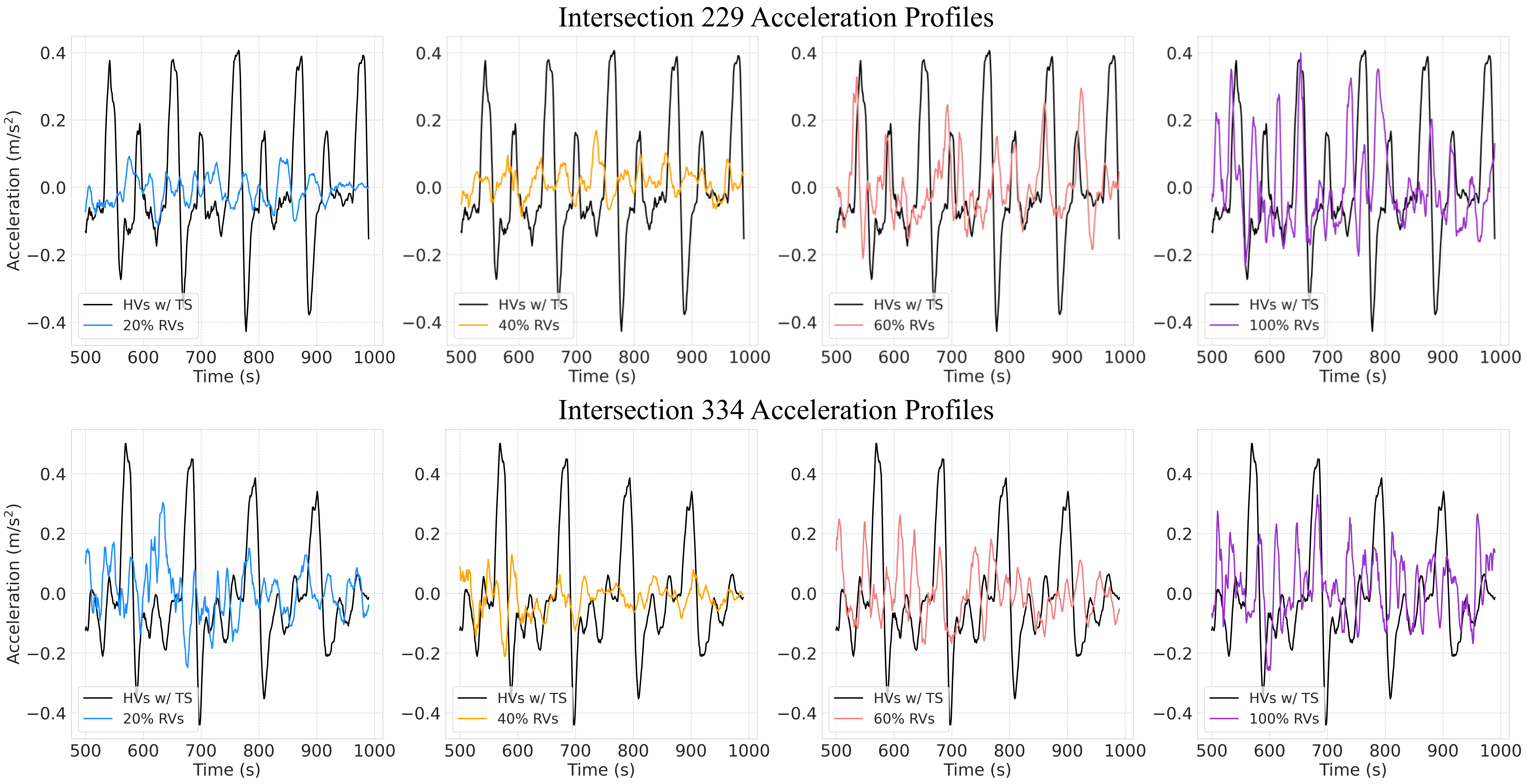}
    \caption{Acceleration profile comparisons between HVs with traffic signals and either 20\%, 40\%, 60\%, or 100\% RV penetration rates at Intersection 229 and Intersection 334. Intersection 229 is selected due to it having a different topology from the other intersections, and Intersection 334 is chosen as a representative for Intersections 499 and 332 due to all three having similar topologies. Each acceleration presented is the average across all vehicles present at the intersections for each second. Values have been smoothed using a moving average with window size of $10$. At 20\%, 40\%, 60\% RV penetration rates for both intersections, we observe more stable acceleration rates around 0 compared to HVs with traffic signals that see periods of sudden, high accelerations. These more stable accelerations results in, on average, less fuel consumption being used. For Intersection 229 at 100\% RV penetration rate, we observe spikes in accelerations similar to the baseline; however, we believe generated emissions are still less due to less excessive idling at the intersections. For Intersection 334, 100\% RV penetration rate sees more accelerations around 0 compared to at Intersection 229, leading to less generated emissions. Overall, the RVs at each considered RV penetration rate exhibits behavior that can lead to reduced emissions.}
    \label{fig:accel_profiles}
    \vspace{-0.5em}
\end{figure*}

To gain further insight as to why the RVs' mixed traffic control strategies lead to fewer emissions, we explore acceleration profiles, Fig.~\ref{fig:accel_profiles}, at 20\%, 40\%, 60\%, and 100\% RV penetration rates at Intersections 229 and 334. We select those four RV penetration rates as they represent key points along the emissions curve from Fig.~\ref{fig:avg_int_results}. We select Intersection 229 as it has a different topology compared to Intersections 499, 332, and 334; we select Intersection 334 as a representative for 499 and 332 as its topology is similar to those two intersections. We consider time (seconds; x-axis) versus acceleration (m/s$^2$; y-axis) where the acceleration values are all vehicle averages at the intersections. We smooth values using moving averages with a window size of $10$.

From the acceleration profiles, we see a general trend at 20\%, 40\%, and 60\% RV penetration rates where the accelerations remain stable near 0 for Intersections 229 and 334. This means there are less severe accelerations to generate emissions. We believe this is partly why even at lower RV penetration rates, generated emissions are lower than with HVs with signalized intersections. At 100\% RV penetration rate for Intersection 229, we observe acceleration spikes similar to HVs at signalized intersections; however, we believe 100\% RV penetration rates generate less emissions compared to the baseline due to a significant decrease in wait time at Intersection 229. This means there is less congestion, and excessive idling that causes emissions. At 100\% RV penetration rate for Intersection 334, we see more stable accelerations around 0, leading to lower generated emissions compared to HVs with signalized intersections. In general, the RVs exhibit acceleration profiles leading to reduced emissions.

\subsubsection{Overall Network Results}

\begin{table*}
    \centering
    \begin{tabular}{lccccccccccc}
        \toprule
         & & \multicolumn{10}{c}{RV Penetration Rate} \\
         \cmidrule(lr){3-12}
         Metric & HVs w/ TS & 10\% & 20\% & 30\% & 40\% & 50\% & \textbf{60\%} & 70\% & \textbf{80\%} & \textbf{90\%} & \textbf{100\%} \\
         \midrule
         Fuel Consumption (ml/s) & 0.8568 & 0.8328 & 0.8296 & 0.7884 & 0.7861 & 0.7977 & 0.7653 & 0.7758 & 0.7696 & 0.7546 & 0.7370 \\
         CO$_2$ Emissions (mg/s) & 1993 & 1937 & 1930 & 1834 & 1829 & 1856 & 1780 & 1805 & 1790 & 1755 & 1715 \\
         CO Emissions (mg/s) & 71.83 & 88.64 & 88.99 & 78.10 & 76.65 & 75.46 & 69.52 & 71.99 & 70.40 & 63.77 & 58.28 \\
         HC Emissions (mg/s) & 0.3732 & 0.4486 & 0.4501 & 0.3982 & 0.3914 & 0.3866 & 0.3578 & 0.3696 & 0.3620 & 0.3310 & 0.3050 \\
         NOx Emissions (mg/s) & 0.8395 & 0.8411 & 0.8389 & 0.7882 & 0.7844 & 0.7924 & 0.7572 & 0.7685 & 0.7609 & 0.7386 & 0.7157 \\
         \bottomrule
    \end{tabular}
    \caption{Results for fuel consumption and CO$_2$, CO, HC, and NOx emissions collected across the entire road network for HVs with traffic signals and RV penetration rates. Bold penetration rates indicate the RVs employed control strategies that outperform HVs with signalized intersections for all emissions. The RVs' mixed traffic control strategies are employed at the unsignalized intersections; however, because traffic flows amongst the unsignalized intersections, we investigate the ripple effects of the RVs' actions. We observe fuel consumption and CO$_2$ emissions are reduced compared to the baseline at every RV penetration rate. However, CO emissions require at least 60\% RV penetration rate, HC emissions require  at least 60\% RV penetration rate, and NOx emissions require at least 20\% RV penetration rates to see improvements. The greatest reduction in generated emissions is achieved at 100\% RV penetration compared to the baseline. In general, the control strategies employed by the RVs, despite being implemented only at the intersections, have a net positive ripple effect on the network's generated emissions.}
    \label{tab:overall_network}
    \vspace{-2.0em}
\end{table*}

Table~\ref{tab:overall_network} presents our results in analyzing the emissions effects on the entire road network from RVs implementing their mixed traffic control strategies. We consider fuel consumption (ml/s), CO$_2$ (mg/s), CO (mg/s), HC (mg/s), and NOx (mg/s). Bold penetration rates mean the RVs outperform HVs at signalized intersections for every emission type. The mixed traffic control strategies employed are only carried out at the individual intersections, with each intersection having their own real-world traffic demand the RVs must learn on; however, the RVs' actions cause ripple effects throughout the road network as traffic flows amongst the intersections. Thus, we investigate the outcomes of said ripple effects on generated emissions. 

From Table~\ref{tab:overall_network}, we observe at every RV penetration rate, fuel consumption and CO$_2$ emissions are less compared to HVs with signalized intersections. For CO emissions and HC emissions, at least 60\% RV penetration rate is required to outperform the baseline (for CO emissions, RVs at 70\% RV penetration rate attain 71.99 mg/s, which is slightly higher than the baseline at 71.83 mg/s). For NOx emissions, with at least 20\% RV penetration rate, we observe RVs generating less emissions compared to the baseline. At 100\% RV penetration rate, we observe the biggest improvements in generated emissions at 14\%, 14\%, 19\%, 18\%, and 15\% for fuel consumption, CO$_2$, CO, HC, and NOx emissions, respectively. Overall, the RVs' control strategies reduce generated emissions across the entire road network compared to HVs with signalized intersections; this means the ripple effects from RVs controlling and coordinating mixed traffic at unsignalized intersections can positively impact the road network as a whole even if the control strategies are only enacted at the intersections.

%% file: sections/conclusion.tex
\section{CONCLUSION and FUTURE WORK}

In this work, we provide an emissions analysis of mixed traffic control at unsignalized intersections. We analyze intersections with complex, real-word topologies with real-world traffic demands from Colorado Springs, CO, USA. We find that fuel consumption, CO$_2$ emissions, and NOx emissions are reduced at every RV penetration by up to 27\%, 27\%, and 28\%, respectively, and CO and HC emissions are reduced with at least 30\% RV penetration rate by up to 42\% and 43\%, respectively, averaged across the four intersections. We also find the RVs tend to stabilize their accelerations near 0, creating less significant generated emissions events. Lastly, we observe RVs employing mixed traffic control strategies at unsignalized intersections has a net benefit effect on the whole road network's generated emissions; we see improvements of up to 19\% on generated emissions.

In future, we plan to go beyond an emissions analysis of mixed traffic control strategies at unsignalized intersections and intentionally develop eco-driving strategies at unsignalized intersections. We intend on accomplishing this by adding a sustainability term to the reward function in Eq.~\ref{reward_function} and changing the action space to allow the robot vehicles to accelerate/decelerate from a continuous range. This gives the robot vehicles higher fidelity control over their behavior, allowing them to more precisely control mixed traffic and generated emissions.
Additionally, we plan to investigate multi-intersection communication of state information to see if such information allows the RVs further improvements to their control strategies and generated emissions. We also plan to research the ability of RVs trained under specific RV penetration rates to generalize to lower penetration rates; should the RVs generalize well, then the RVs will still effectively control mixed traffic in low RV penetration rate scenarios.

%% file: sections/acknowledge.tex
\section*{ACKNOWLEDGEMENT}
This research is supported by NSF IIS-2153426. The authors would also like to thank NVIDIA and the Center for Transportation Research (CTR) at the University of Tennessee, Knoxville for their support.